\documentclass[12pt]{article}
\usepackage{epsfig, epsf, graphicx, subfigure}
\usepackage{pstricks, pst-node, psfrag}
\usepackage{amssymb,amsmath,bm}
\usepackage{verbatim,enumerate}
\usepackage{setspace}
\usepackage{enumitem}
\usepackage{array,epsfig,fancyheadings,rotating}
\usepackage[]{hyperref}  

\usepackage{tabularx,ragged2e,booktabs,caption}
\usepackage[english]{babel}
\usepackage[sectionbib]{natbib}
\usepackage{multirow}
\usepackage{theorem}
\usepackage{float}
\def\00{\mathrm{0}}

\newtheorem{thm}{Theorem}
\newtheorem{defn}{Definition}

\begin{document}

\thispagestyle{empty} \baselineskip=28pt \vskip 5mm
\begin{center} {\Large{\bf An Outlyingness Matrix for Multivariate Functional Data Classification}}
\end{center}

\baselineskip=12pt \vskip 10mm

\begin{center}\large
Wenlin Dai and Marc G. Genton\footnote[1]{
\baselineskip=10pt CEMSE Division,
King Abdullah University of Science and Technology,
Thuwal 23955-6900, Saudi Arabia. E-mail: wenlin.dai@kaust.edu.sa, marc.genton@kaust.edu.sa\\
This research was supported by the
King Abdullah University of Science and Technology (KAUST).}
\end{center}

\baselineskip=17pt \vskip 10mm \centerline{\today} \vskip 15mm

\begin{center}
{\large{\bf Abstract}}
\end{center}

The classification of multivariate functional data is an important task in scientific research.
Unlike point-wise data, functional data are usually classified by their shapes rather than by their scales.
We define an outlyingness matrix by extending directional outlyingness, an effective measure of the shape variation of curves that combines the direction of outlyingness with conventional statistical depth.
We propose classifiers based on directional outlyingness and the outlyingness matrix.
Our classifiers provide better performance compared with existing depth-based classifiers when applied on both univariate and multivariate functional data from simulation studies.
We also test our methods on two data problems: speech recognition and gesture classification, and obtain results that are consistent with the findings from the simulated data.

\baselineskip=14pt

\par\vfill\noindent

\clearpage\pagebreak\newpage \pagenumbering{arabic}
\baselineskip=26pt

\vskip 24pt
\section{Introduction}
Functional data are frequently collected by researchers in such fields as biology, finance, geology, medicine, and meteorology.
As with other types of data, problems such as ranking, registration, outlier detection, classification, and modeling also arise with functional data.
Many methods have been proposed to extract useful information from functional data (\citet{ramsayfunctional}, \citet{ferraty2006nonparametric}, and  \citet{horvath2012inference}).
Functional classification is an essential task in many applications, e.g., diagnosing diseases based on curves or images from medical test results, recognizing handwriting or speech patterns, and classifying products (\citet{epifanio2012shape}, \citet{delaigle2012achieving}, \citet{alonso2012supervised}, \citet{sguera2014spatial}, and \citet{galeano2015mahalanobis}).

Statistical depth was initially defined to rank multivariate data, mimicking the natural order of univariate data. \citet{zuo2000general} presented details on statistical depth. Recently, the concept has been generalized to functional depth to rank functional data from the center outward (\citet{fraiman2001trimmed}, \citet{cuevas2007robust}, \citet{lopez2009concept}, \citet{lopez2014simplicial}, and \citet{claeskens2014multivariate}).
An alternative way to rank functional data is the tilting approach proposed by \citet{genton2016tilting}.
Functional depth, as a measure of the centrality of curves, has been used extensively to classify functional data, especially if the dataset is possibly contaminated (\citet{sguera2014spatial}).
\citet{lopez2006depth} defined (modified) band depth for functional data, based on which they proposed two methods for classification of functional data: ``distance to the trimmed mean'' and ``weighted averaged distance''.
\citet{cuevas2007robust} introduced random projection depth and the ``within maximum depth'' criterion.
\citet{sguera2014spatial} defined kernelized functional spatial depth and comprehensively investigated the performance of depth-based classifiers.
\citet{cuesta2015dd} and \citet{mosler2015fast} discussed functional versions of the depth-depth (DD) classifier (\citet{li2012dd} and \citet{liu1999multivariate}).
\citet{hubert2016multivariate}
 investigated functional bag distance and a distance-distance plot to classify functional data. \citet{kuhnt2016angle} proposed a graphical approach using the angles in the intersections of one observation with the others.

There have been many other attempts to tackle the challenge of functional data classification, a great number of which sought to generalize finite-dimensional methods to functional settings.
These approaches map functional data to finite-dimensional data via dimension reduction and then apply conventional classification methods, e.g., linear discriminant analysis (LDA) or support vector machines (SVM) (\citet{boser1992training} and \citet{cortes1995support}), to the finite-dimensional data.
Dimension reduction techniques mainly fall into two categories: \emph{regularization} and \emph{filtering}.
The regularization approach treats functional data as multivariate data observed at discrete time points or intervals (\citet{li2008classification} and \citet{delaigle2012componentwise}), and the filtering approach approximates each curve by a linear combination of a finite number of basis functions, representing the data by their corresponding basis coefficients (\citet{james2001functional},  \citet{rossi2006support}, \citet{epifanio2012shape}, \citet{galeano2015mahalanobis}, \citet{li2017functional}, and \citet{yao2016probability}).

Most of the aforementioned methods focus on univariate functional data.
Very little attention has been paid to multivariate functional cases, now frequently observed in scientific research.
Examples of multivariate functional cases are gait data and handwriting data (\citet{ramsayfunctional}), height and weight of children by age (\citet{lopez2014simplicial}) and various records from weather stations (\citet{claeskens2014multivariate}).
Classifying such multivariate functional data jointly rather than marginally is necessary because a joint method takes into consideration the interaction between components and one observation may be marginally assigned to different classes by different components.

Locations/coordinates are used to classify point-wise data; however, the variation between different groups of curves in functional data classification usually results from the data's different patterns/shapes rather than their scales. We refer the readers to the simulation settings and applications in a number of references (\citet{cuevas2007robust}, \citet{alonso2012supervised}, \citet{epifanio2012shape}, \citet{galeano2015mahalanobis}, and \citet{sguera2014spatial}).
This important feature of functional data classification cannot be handled by conventional functional depths which do not effectively describe the differences in shapes of curves.
A recently proposed notion of directional outlyingness (\citet{dai2017directional}) overcomes these drawbacks.
The authors pointed out that the direction of outlyingness is crucial to describing the centrality of multivariate functional data.
By combining the direction of outlyingness with the conventional point-wise outlyingness, they established a framework that can decompose total functional outlyingness into shape outlyingness and scale outlyingness.
The shape outlyingness measures the change of point-wise outlyingness in view of both level and direction.
It thus effectively describes the shape variation between curves.
We extend the scalar outlyingness to an outlyingness matrix, which contains pure information of shape variation of a curve.
Based on directional outlyingness and the outlyingness matrix, we propose two classification methods for multivariate functional data.

The remainder of the paper is organized as follows.
In Section 2, we briefly review the framework of directional outlyingness, define the outlyingness matrix and propose two classification methods for multivariate functional data using this framework.
In Section 3, we evaluate our proposed classifiers on both univariate and multivariate functional data via simulation studies.
In Section 4, we use two datasets to illustrate the performance of the proposed methods in practice.
We end the paper with a short discussion in Section 5. Two illustrative figures of multivariate functional data and proofs for the theoretical results are provided in an online supplement.

\vskip 24pt
\section{Directional Outlyingness and Classification Procedure}
With $K\ge 2$ groups of data as training sets, to classify a new observation from the test set, $\mathbf{X}_0$, into one of the groups, one needs to find an effective measure of distance between $\mathbf{X}_0$ and each groups.
Such a measure is the Bayesian probability for the naive Bayes classifier, the Euclidean distance for the $k$-nearest neighbors classifier, or the functional outlyingness/depth for the depth-based classifier.
Our classification methods fall into the latter category.
In what follows, we first review the framework of directional outlyingness as our measure for the distance between a new curve and a labeled group of curves, and then propose two classification methods based on this framework.

\vskip 12pt
\subsection{Directional Outlyingness}
Consider a $p$-variate stochastic process of continuous functions, $\mathbf{X}=(X_1,\dots,X_p)^{\rm T}$, with each $X_k$ ($1\le k\le p$): $\mathcal{I} \to \mathbb{R}$, $t\mapsto X_k(t)$ from the space $\mathcal{C}(\mathcal{I},\mathbb{R})$ of real continuous functions on $\mathcal{I}$.
At each fixed time point, $t$, $\mathbf{X}(t)$ is a $p$-variate random variable.
Here, $p$ is a finite positive integer that indicates the dimension of the functional data and $\mathcal{I}$ is a compact time interval.
We get univariate functional data when $p=1$ and multivariate functional data when $p\ge2$.
Denote the distribution of $\mathbf{X}$ as $F_{\mathbf{X}}$ and the distribution of $\mathbf{X}(t)$, which is the function value of $\mathbf{X}$ at time point $t$, as $F_{\mathbf{X}(t)}$.
For a sample of curves from $F_{\mathbf{X}}$, $\mathbf{X}_1,\dots,\mathbf{X}_n$, the empirical distribution is denoted as $F_{\mathbf{X},n}$; correspondingly, the empirical distribution of $\mathbf{X}_1(t),\dots,\mathbf{X}_n(t)$ is denoted as $F_{\mathbf{X}(t),n}$. Let $d(\mathbf{X}(t),F_{\mathbf{X}(t)})$: $\mathbb{R}^p \longrightarrow [0,1]$ be a statistical depth function for $\mathbf{X}(t)$ with respect to $F_{\mathbf{X}(t)}$. The finite sample depth function is then denoted as $d_n(\mathbf{X}(t),F_{\mathbf{X}(t),n})$.

Directional outlyingness (\citet{dai2017directional}) is defined by combining conventional statistical depth with the direction of outlyingness. For multivariate point-wise data, assuming $d(\mathbf{X}(t),F_{\mathbf{X}(t)})> 0$, the directional outlyingness is defined as
$$\mathbf{O}(\mathbf{X}(t),F_{\mathbf{X}(t)})=\left\{{1/d(\mathbf{X}(t),F_{\mathbf{X}(t)})}-1\right\}\cdot \mathbf{v}(t),$$
where $\mathbf{v}(t)$ is the unit vector pointing from the median of $F_{\mathbf{X}(t)}$ to $\mathbf{X}(t)$. Specifically, assuming that $\mathbf{Z}(t)$ is the unique median of $F_{\mathbf{X}(t)}$, $\mathbf{v}(t)$ can be expressed as $\mathbf{v}(t)=\left\{\mathbf{X}(t)-\mathbf{Z}(t)\right\}/\|\mathbf{X}(t)-\mathbf{Z}(t)\|$, where $\|\cdot\|$ denotes the $L_2$ norm. Then, \citet{dai2017directional} defined three measures of directional outlyingness for functional data\\
the functional directional outlyingness {\rm ({FO})} is
$${\rm FO}(\mathbf{X},F_{\mathbf{X}})=\int_{\mathcal{I}} \|\mathbf{O}(\mathbf{X}(t),F_{\mathbf{X}(t)})\|^2w(t){\rm d}t;$$
the mean directional outlyingness {\rm (\textbf{MO})} is
$$\mathbf{MO}(\mathbf{X},F_{\mathbf{X}})=\int_{\mathcal{I}} \mathbf{O}(\mathbf{X}(t),F_{\mathbf{X}(t)}) w(t){\rm d}t;$$
the variation of directional outlyingness {\rm (VO)} is
$${\rm VO}(\mathbf{X},F_{\mathbf{X}})=\int_{\mathcal{I}} \|\mathbf{O}(\mathbf{X}(t),F_{\mathbf{X}(t)})-{\rm \mathbf{MO}}(\mathbf{X},F_{\mathbf{X}})\|^2w(t){\rm d}t,$$
where $w(t)$ is a weight function defined on $\mathcal{I}$, which can be constant or proportional to the local variation at each time point (\citet{claeskens2014multivariate}).
Throughout, we use a constant weight function, $w(t)=\{\lambda(\mathcal{I})\}^{-1}$, where $\lambda(\cdot)$ is Lebesgue measure.
$\mathbf{MO}$ indicates the position of a curve relative to the center on average, which measures the scale outlyingness of this curve; ${\rm VO}$ represents the variation in the quantitative and directional aspects of the directional outlyingness of a curve and measures the shape outlyingness of that curve.
We can link the three measures of directional outlyingness by
\begin{eqnarray}
{\rm FO}(\mathbf{X},F_{\mathbf{X}})=\|\mathbf{MO}(\mathbf{X},F_{\mathbf{X}})\|^2+{\rm VO}(\mathbf{X},F_{\mathbf{X}}). \label{decomp_scaler}
\end{eqnarray}
Then, ${\rm FO}$ can be regarded as the overall outlyingness and is equivalent to the conventional functional outlyingness. When the curves are parallel to each other, ${\rm VO}$ is zero and a quadratic relationship exists between ${\rm FO}$ and $\mathbf{MO}$.
Many existing statistical depths can be used to construct their corresponding directional outlyingness, among which we suggest the distance-based depths, e.g., random projection depth (\citet{zuo2003projection}) and the Mahalanobis depth (\citet{zuo2000general}). In the current paper, we choose the Mahalanobis depth to construct directional outlyingness for all the numerical studies. As an intuitive illustration of this framework, an example is provided in the supplement.

Compared with conventional functional depths, directional outlyingness  more effectively describes the centrality of functional data, especially the shape variation, because ${\rm VO}$ accounts for not only variation of absolute values of point-wise outlyingness but also for the change in their directions.
This advantage coincides with the functional data classification task, which is essentially to distinguish curves by their differences in shapes rather than scales. 
With the above advantages, we adopt the functional directional outlyingness to measure the distance between the curve to be classified and the labeled groups of curves.
In the next two subsections, we propose two classification methods for multivariate functional data.
Both are based on a similar idea used by the maximum depth classifier: a new curve should be assigned to the class leading to the smallest outlyingness value.


\vskip 12pt
\subsection{Two-Step Outlyingness}
In the first step, directional outlyingness maps one $p$-variate curve to a $(p+1)$-dimensional vector, $\mathbf{Y}=(\mathbf{MO}^{\rm T},{\rm VO})^{\rm T}$, that involves both magnitude outlyingness and shape outlyingness of this curve.
As shown in Figure S1 of the supplement, the $\mathbf{Y}_i$'s that correspond to the outlying curves are also isolated from the cluster of points corresponding to non-outlying curves.
In the second step, we can simply measure the outlyingness of the point, $\mathbf{Y}_i$, to assess the outlyingness of its respective curve, $\mathbf{X}_i$.
Specifically, we calculate the Mahalanobis distance (\citet{mahalanobis1936generalized}) of $\mathbf{Y}_i$ and employ this distance as a two-step outlyingness of the raw curve.


For a set of $n$ observations, $\mathbf{Y}_i$ ($i=1,\dots,n$), a general form of the Mahalanobis distance is $${\rm D}(\mathbf{Y},\pmb{\mu})=\sqrt{(\mathbf{Y}-\pmb{\mu})^{\rm T}\mathbf{S}^{-1}(\mathbf{Y}-\pmb{\mu})},$$
where $\pmb{\mu}$ is the mean vector of the $\mathbf{Y}_i$'s and $\mathbf{S}$ is the covariance matrix.
Various estimators of $\mathbf{S}$ exist in the literature, among which the minimum covariance determinant (MCD) estimator (\citet{rousseeuw1985multivariate}) is quite popular due to its robustness.
To subtract the influence of potential outliers, we utilize this estimator to calculate the distance for our method.

In particular, the robust Mahalanobis distance based on MCD and a sample of size $h\le n$ can be expressed as
$${\rm RMD}_{\rm J}(\mathbf{Y})=\sqrt{(\mathbf{Y}-\mathbf{\bar Y^*}_{\rm J})^{\rm T}{\mathbf{S}^*_{\rm J}}^{-1}(\mathbf{Y}-\mathbf{\bar Y^*}_{\rm J})},$$
where $\rm J$ denotes the set of $h$ points that minimizes the determinant of the corresponding covariance matrix, $\mathbf{\bar Y^*}_{\rm J}=h^{-1}\sum_{i\in \rm J}\mathbf{Y}_{i}$ and
$\mathbf{S}^*_{\rm J}=h^{-1}\sum_{i\in \rm J}(\mathbf{Y}_i-\mathbf{\bar Y^*}_{\rm J})(\mathbf{Y}_i-\mathbf{\bar Y^*}_{\rm J})^{\rm T}$. The sub-sample size, $h$, controls the robustness of the method.
For a $(p+1)$-dimensional distribution, the maximum finite sample breakdown point is $[(n-p)/2]/n$, where $[a]$ denotes the integer part of $a\in \mathbb{R}$.
Assume that we have $K\ge 2$ groups of functional observations, $G_i$ ($i=1,\dots,K$). To classify a new curve, $\mathbf{X}_0$, into one of the groups, we use the classifier
$$C_1={\rm arg}\min_{1\le i\le K}\left\{{\rm RMD}_{G_i}(\mathbf{X}_0)\right\},$$
where $C_1$ is the group label, to which we assign $\mathbf{X}_0$, and ${\rm RMD}_{G_i}(\mathbf{X}_0)$ is the robust Mahalanobis distance of $\mathbf{X}_0$ to $G_i$.
This classifier is based on an idea similar to the ``within maximum depth'' criterion (\citet{cuevas2007robust}) that assigns a new observation to the group that leads to a larger depth.
The difference is that we use a two-step outlyingness, which can better distinguish shape variation between curves compared with conventional functional depths utilized in existing methods.

\vskip 12pt
\subsection{Outlyingness Matrix}
Unlike conventional statistical depth, point-wise directional outlyingness of multivariate functional data, $\mathbf{O}(\mathbf{X}(t),F_{\mathbf{X}(t)})$, is a vector
that allows us to define two additional statistics to describe the centrality of multivariate functional data.

\vskip 5pt
{\defn {\rm (\textbf{Outlyingness Matrix of Multivariate Functional Data}):} \label{Definition 1}
Consider a stochastic process, $\mathbf{X}: \mathcal{I} \longrightarrow \mathbb{R}^p$, that takes values in the space $\mathcal{C}(\mathcal{I},\mathbb{R}^p)$ of real continuous functions defined from a compact interval,
$\mathcal{I}$, to $\mathbb{R}^p$ with probability distribution $F_{\mathbf{X}}$. The
functional directional outlyingness matrix {\rm (\textbf{FOM})} is
$${\rm \mathbf{FOM}}(\mathbf{X},F_{\mathbf{X}})=\int_{\mathcal{I}} \mathbf{O}(\mathbf{X}(t),F_{\mathbf{X}(t)})\mathbf{O}^{\rm T}(\mathbf{X}(t),F_{\mathbf{X}(t)})w(t){\rm d}t;$$
and the variation of directional outlyingness matrix {\rm (\textbf{VOM})} is}
$${\rm \mathbf{VOM}}(\mathbf{X},F_{\mathbf{X}})=\int_{\mathcal{I}} \left\{\mathbf{O}(\mathbf{X}(t),F_{\mathbf{X}(t)})-{\rm \mathbf{MO}}(\mathbf{X},F_{\mathbf{X}})\right\}\left\{\mathbf{O}(\mathbf{X}(t),F_{\mathbf{X}(t)})-{\rm \mathbf{MO}}(\mathbf{X},F_{\mathbf{X}})\right\}^{\rm T}w(t){\rm d}t.$$
\vskip 5pt
\noindent
${\rm \mathbf{FOM}}$ can be regarded as a matrix version of the total outlyingness, ${\rm FO}$, and ${\rm \mathbf{VOM}}$ corresponds to the shape outlyingness, ${\rm VO}$.
A decomposition of ${\rm \mathbf{FOM}}$ and its connection with the scalar statistics are on exhibit in the following.

\vskip 5pt
{\thm {\rm (\textbf{Outlyingness Decomposition}):}\label{Theorem 1} For the statistics in Definition \ref{Definition 1}, we have
\begin{itemize}[noitemsep]
\item [\rm (i)] ${\rm \mathbf{FOM}}(\mathbf{X},F_{\mathbf{X}})=\mathbf{MO}(\mathbf{X},F_{\mathbf{X}})\mathbf{MO}^{\rm T}(\mathbf{X},F_{\mathbf{X}})+{\rm \mathbf{VOM}}(\mathbf{X},F_{\mathbf{X}})$;
\item [\rm (ii)] ${\rm FO}(\mathbf{X},F_{\mathbf{X}})={\rm tr}\left\{{\rm \mathbf{FOM}}(\mathbf{X},F_{\mathbf{X}})\right\}$ and ${\rm VO}(\mathbf{X},F_{\mathbf{X}})={\rm tr}\left\{{\rm \mathbf{VOM}}(\mathbf{X},F_{\mathbf{X}})\right\}$, where ${\rm tr}(\cdot)$ denotes the trace of a matrix.
\end{itemize}
}

\vskip 5pt
{\thm \label{Theorem 2} {\rm (\textbf{Properties of the Outlyingness Matrix}):} Assume that $\mathbf{O}\left(\mathbf{X}(t), F_{\mathbf{X}(t)}\right)$ is a valid directional outlyingness for point-wise data from \citet{dai2017directional}. Then, for a constant weight function, we have
\begin{eqnarray*}
\mathbf{VOM}\left(\mathbf{T}(\mathbf{X}_g),F_{\mathbf{T}(\mathbf{X}_g)}\right)=\mathbf{A}_0\mathbf{VOM}\left(\mathbf{X},F_{\mathbf{X}}\right){\mathbf{A}_0}^{\rm T},
\end{eqnarray*}
where $\mathbf{T}(\mathbf{X}_g(t))=\mathbf{A}\left\{g(t)\right\}\mathbf{X}\left\{g(t)\right\}+\mathbf{b}\left\{g(t)\right\}$ is a transformation of $\mathbf{X}$ in both the response and support domains, $\mathbf{A}(t)=f(t){\mathbf{A}_0}$ with $f(t)>0$ for $t\in \mathcal{I}$ and $\mathbf{A}_0$ an orthogonal matrix, $\mathbf{b}(t)$ is an $p$-vector at each time $t$, and $g$ is a bijection on the interval $\mathcal{I}$.
}

\noindent
W focus on the cases when the distinction between different groups of functional data depends on their patterns/shapes.
$\mathbf{VOM}$ effectively measures the level of shape variation between one curve and a group of curves. Our second classifier is
$$C_2={\rm arg}\min_{1\le i\le K}\left\{\|{\mathbf{VOM}}(\mathbf{X}_0,G_i)\|_{F}\right\},$$
where $\|\cdot\|_F$ denotes the Frobenius norm of a matrix and $C_2$ is the group label, to which we assign $\mathbf{X}_0$.
Compared with our first classifier, this classifier is based purely on the shape information.
We choose the Frobenius norm to get a scalar to take into consideration the interaction between outlyingness in different directions (the off-diagonal elements of $\mathbf{VOM}$).

\vskip 24pt
\section{Simulation Studies}
In this section, we report on some simulation studies to assess finite-sample performances of the proposed classification methods and to compare them with those of some existing methods based on conventional statistical depth.
We investigate both univariate and multivariate functional data cases.
\subsection{Classification Methods}
We calculated the point-wise directional outlyingness with the Mahalanobis depth (MD) (\citet{zuo2000general}) for our proposed methods, two-step outlyingness, denoted by RMD, and outlyingness matrix, denoted by $\mathbf{VOM}$.
We considered the ``within maximum depth'' criterion (\citet{cuevas2007robust}) for existing methods, using four conventional functional depths that can handle both univariate and multivariate functional data.
\begin{description}
 \item[Method FM1.] Integrated depth defined by \citet{fraiman2001trimmed}, which calculates functional depth as the integral of point-wise depth across the whole support interval of a curve. We used random Tukey depth (TD) (\citet{tukey1975mathematics}) as the point-wise depth for this method.
 \item[Method FM2.] Integrated depth with MD as the point-wise depth.
 The R functions \emph{depth.FM} and \emph{depth.FMp} in the package \emph{fda.usc} were used to calculate FM1 and FM2 for univariate and multivariate cases, respectively.
 \item[Method RP1.] Random projection depth defined by \citet{cuevas2007robust}. In this method, we randomly chose ${\rm NR}$ directions, projected the curves onto each direction, calculated the statistical depth based on the projections for each direction and took the average of the direction-wise depth. Here we set the number of random directions, ${\rm NR}=50$. The direction in this method refers to a random function, $a$, in the Hilbert space $L^2[0,1]$ so that the projection of a datum, $X$, is given by the standard inner product $\langle a,X\rangle=\int_{0}^1 a(t)X(t){\rm d}t$. We used TD as the direction-wise depth for this method.
 \item[Method RP2.] Random projection depth with MD as the direction-wise depth.  The R functions \emph{depth.RP} and \emph{depth.RPp} in the package \emph{fda.usc} were used to calculate RP1 and RP2 for univariate and multivariate cases, respectively.
\end{description}
TD and MD were selected as representatives of rank-based and distance-based depths, respectively.
Aside from them, many other notions have been proposed in the literature.
Some methods can be regarded as special cases of FM1 (with different point-wise depths), including modified band depth (\citet{lopez2009concept}), half-region depth (\citet{lopez2011half}), simplicial band depth (\citet{lopez2014simplicial}), multivariate functional halfspace depth (\citet{claeskens2014multivariate}), and multivariate functional skew-adjusted projection depth (\citet{hubert2015multivariate}).
Some methods have been specifically designed for univariate functional data, including kernelized functional spatial depth (\citet{sguera2014spatial}) and extremal depth (\citet{narisetty2016extremal}).

\vskip 12pt
\subsection{Univariate Functional Data}
We considered three univariate settings. Different groups of curves vary in terms of patterns or shapes rather than scales. Each pair of curves thus oscillates within a similar range in different fashions in our settings.

\vskip 5pt
\noindent
\textbf{Data 1}. Class 0: $X_0(t)=u_{01}\sin(2\pi t)+u_{02}\cos(2 \pi t)+\varepsilon(t)$ and class 1: $X_1(t)=u_{11}\sin(2\pi t)+u_{12}\cos(2 \pi t)+\varepsilon(t)$, where $u_{01}$ and $u_{02}$ were generated independently from a uniform distribution $U(0.5,1)$, $u_{11}$ and $u_{12}$ were i.i.d. observations from $U(1,1.2)$ and
$\varepsilon(t)$ was a Gaussian process with covariance function
$${\rm cov}\{\varepsilon(t),\varepsilon(s)\}=0.25\exp\{-(t-s)^2\}, \quad t~,s\in [0,1].$$
This setting has been considered by \citet{sguera2014spatial}.

\vskip 5pt
\noindent
\textbf{Data 2}. Class 0: $X_0(t)=10\sin(2\pi t)+\varepsilon(t)$ and class 1: $X_1(t)=10\sin(2\pi t)+\sin(20\pi t)+\varepsilon(t)$. A similar setting has been considered by \citet{cuevas2007robust}.

\vskip 5pt
\noindent
\textbf{Data 3}. Class 0: $X_0(t)=u_0\sin(2\pi t)+\varepsilon(t)$ and class 1: $X_1(t)=u_1+\varepsilon(t)$, where $u_0$ was generated from $U(0.5,1)$ and $u_1$ was generated from $U(-1,1)$. \citet{lopez2009concept} considered a similar setting for outlier detection.

\begin{figure}[t!]
\begin{center}
\includegraphics[width=16.5cm,height=12cm]{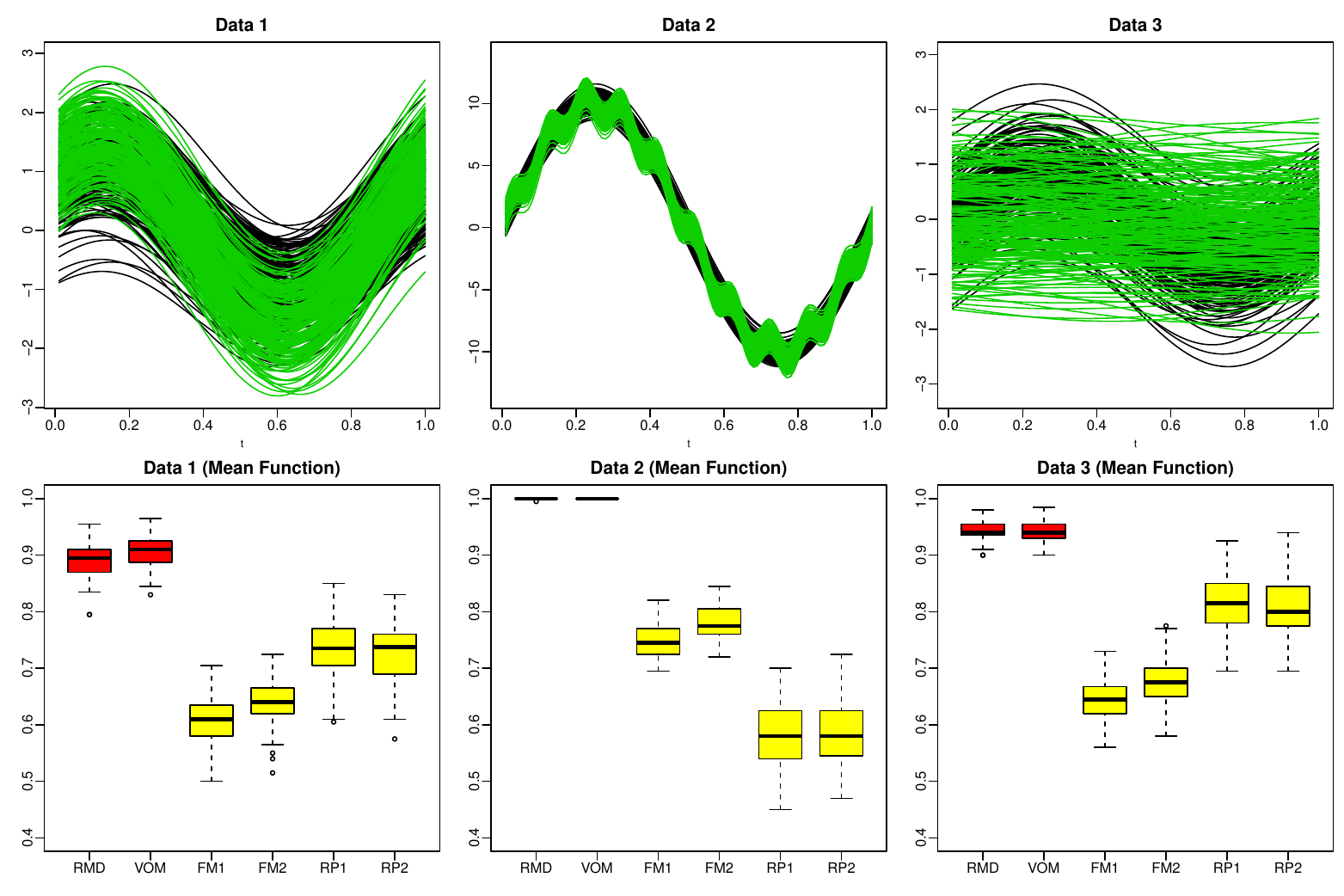}\\
\caption{Top panel: Realizations of three univariate functional data settings (Data 1, 2, 3) with two classes. Bottom panel: correct classification rates of
our two proposed methods, RMD and VOM, and four existing methods, FM1, FM2, RP1, and RP2, for three settings based on 100 simulations.}
\label{plot_univariate}
\end{center}
\end{figure}

In the top panel of Figure \ref{plot_univariate}, we provide one realization of two classes of curves for each setting. The functions were evaluated at 50 equidistant points on $[0,1]$.
We independently generated 200 samples from both classes of each data setting, randomly chose 100 of them as the training set, and treated the remaining 100 samples as the testing set.
We applied the six methods to the generated data and calculated the correct classification rate, $p_c$, for each method.
We repeated the above procedure 100 times. The results are presented in the bottom panel of Figure \ref{plot_univariate}.
Under all three settings, our proposed methods performed significantly better than the four existing classification methods.
For example, the classification result from our methods are almost perfect, whereas the other four methods achieve $p_c$ less than $80\%$ in the second setting, because our methods describe the shape variation of a curve more effectively than does conventional functional depth.

\vskip 12pt
\subsection{Multivariate Functional Data}
Typically, multivariate functional data are obtained from two sources: combining raw univariate curves and their derivatives (\citet{cuevas2007robust} and \citet{claeskens2014multivariate}) or functional data with multiple responses (\citet{hubert2016multivariate} and \citet{hubert2015multivariate}). We conducted simulation studies on both sources.

\begin{figure}[t!]
\begin{center}
\includegraphics[width=16.5cm,height=6cm]{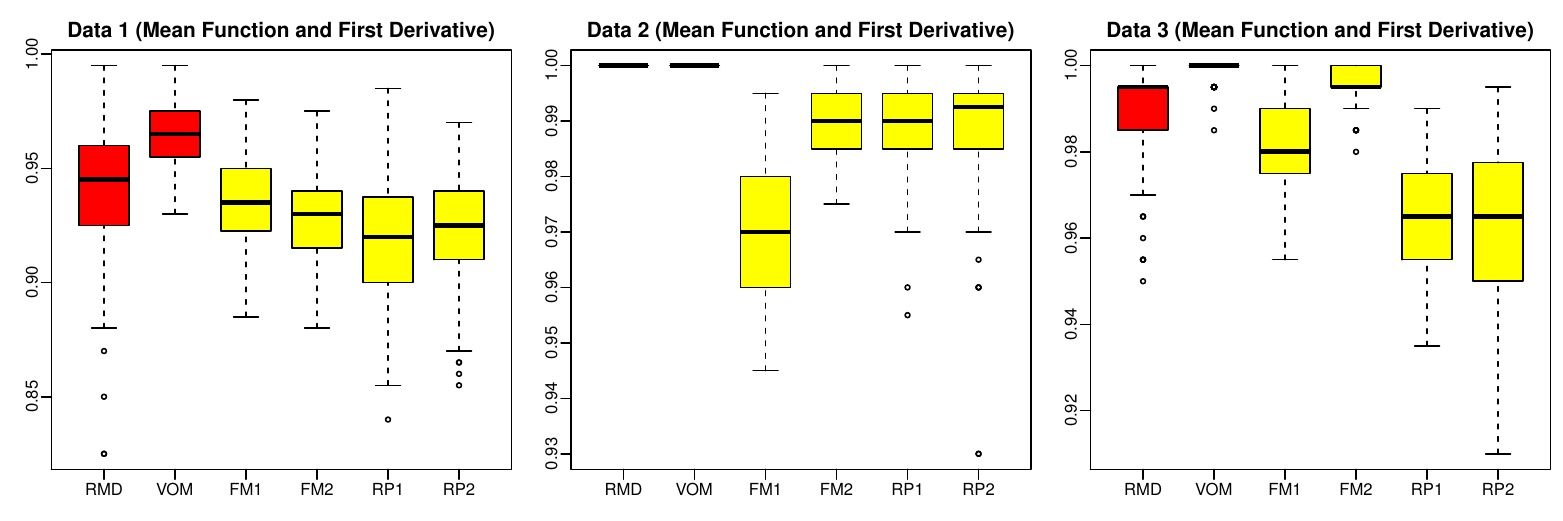}\\
\caption{Correct classification rates of
our two proposed methods, RMD and VOM, and four existing methods, FM1, FM2, RP1, and RP2, for three settings (Data 1, 2, 3) based on 100 simulations, using both mean functions and first-order derivatives.}
\label{univariate_derivative_result}
\end{center}
\end{figure}

In the first scenario, we combined mean functions and the first-order derivatives of Data 1, 2, and 3 to get bivariate functional data.
Under the same setting for sample sizes, design points, and repeated times, we applied the six methods to the resulting data and present the classification results in Figure \ref{univariate_derivative_result}.
On the three datasets, RMD and VOM perform better than the existing methods and VOM always performs the best. The performance of the existing methods improves by combining the first-order derivatives with the mean function for classification; the derivatives are no longer of the same scale for different groups, which makes classifying by conventional functional depths easier.

In the second scenario, we considered three settings: two bivariate cases and one three-variate case. Again, the two classes of simulated data possess the same range but different patterns.
\vskip 5pt
\noindent
\textbf{Data 4}. Class 0: $\mathbf{X}_0=(X_{01},X_{02})^{\rm T}$ with $X_{01}(t)=\sin(4\pi t)+e_1(t)~{\rm and}~X_{02}(t)=\cos(4\pi t)+e_2(t)$ and class 1: $\mathbf{X}_1=(X_{11},X_{12})^{\rm T}$ with $X_{11}(t)=\sin(4\pi t)+\sin(20\pi t)/10+e_1(t)~{\rm and}~X_{12}(t)=\cos(4\pi t)+\cos(20\pi t)/10+e_2(t)$, where $\textbf{e}(t)=\{e_1(t),e_2(t)\}^{\rm T}$ was a bivariate Gaussian process with zero mean and covariance function (\citet{gneiting2010matern} and \citet{apanasovich2012valid}):
$${\rm cov}\{e_i(s),e_j(t)\}=\rho_{ij}\sigma_{i}\sigma_{j}\mathcal{M}(|s-t|;\nu_{ij},\alpha_{ij}), \quad i,j=1,2,$$
where $\rho_{12}$ is the correlation between $X_{i1}(t)$ and $X_{i2}(t)$ ($i=0,1$), $\rho_{11}=\rho_{22}=1$, $\sigma_i^2$ is the marginal variance, and $\mathcal{M}(h;\nu,\alpha)=2^{1-\nu}\Gamma(\nu)^{-1}\left(\alpha|h|\right)^{\nu}\mathcal{K}_\nu(\alpha|h|)$ with $|h|=|s-t|$ is the Mat{\'e}rn class (\citet{matern1960spatial}) where $\mathcal{K}_\nu$ is a modified Bessel function of the second kind of order $\nu$, $\nu>0$ is a smoothness parameter, and $\alpha>0$ is a range parameter. Here, we set $\sigma_1=\sigma_2=0.01$, $\nu_{11}=\nu_{22}=\nu_{12}=2$, $\alpha_{11}=0.2$, $\alpha_{22}=0.1$, $\alpha_{12}=0.16$, and $\rho_{12}=0.6$.

\vskip 5pt
\noindent
\textbf{Data 5}. Class 0: $\mathbf{X}_0=(X_{01},X_{02})^{\rm T}$ with $X_{01}(t)=U_{01}+e_1(t)~{\rm and}~X_{02}(t)=U_{02}+e_2(t)$ and class 1: $\mathbf{X}_1=(X_{11},X_{12})^{\rm T}$ with $X_{11}(t)=U_{11}+\sin(4\pi t)+e_1(t)~{\rm and}~X_{12}(t)=U_{12}+\cos(4\pi t)+e_2(t)$, where $U_{01}$ were generated independently from $U(-1.5,1.5)$, $U_{01}$ and $U_{02}$ were generated independently from $U(-2,2)$; $U_{11}$ and $U_{12}$ were generated independently from $U(-0.5,0.5)$.

\vskip 5pt
\noindent
\textbf{Data 6}. Class 0: $\mathbf{X}_0=(X_{01},X_{02},X_{03})^{\rm T}$ with three components generated from class 0 of Data 1, 2, and 3.
Class 1: $\mathbf{X}_1=(X_{11},X_{12},X_{13})^{\rm T}$ with three components generated from class 1 of Data 1, 2, and 3. Data 6 is a three-variate setting.

\begin{figure}[t!]
\begin{center}
\includegraphics[width=16.5cm,height=12cm]{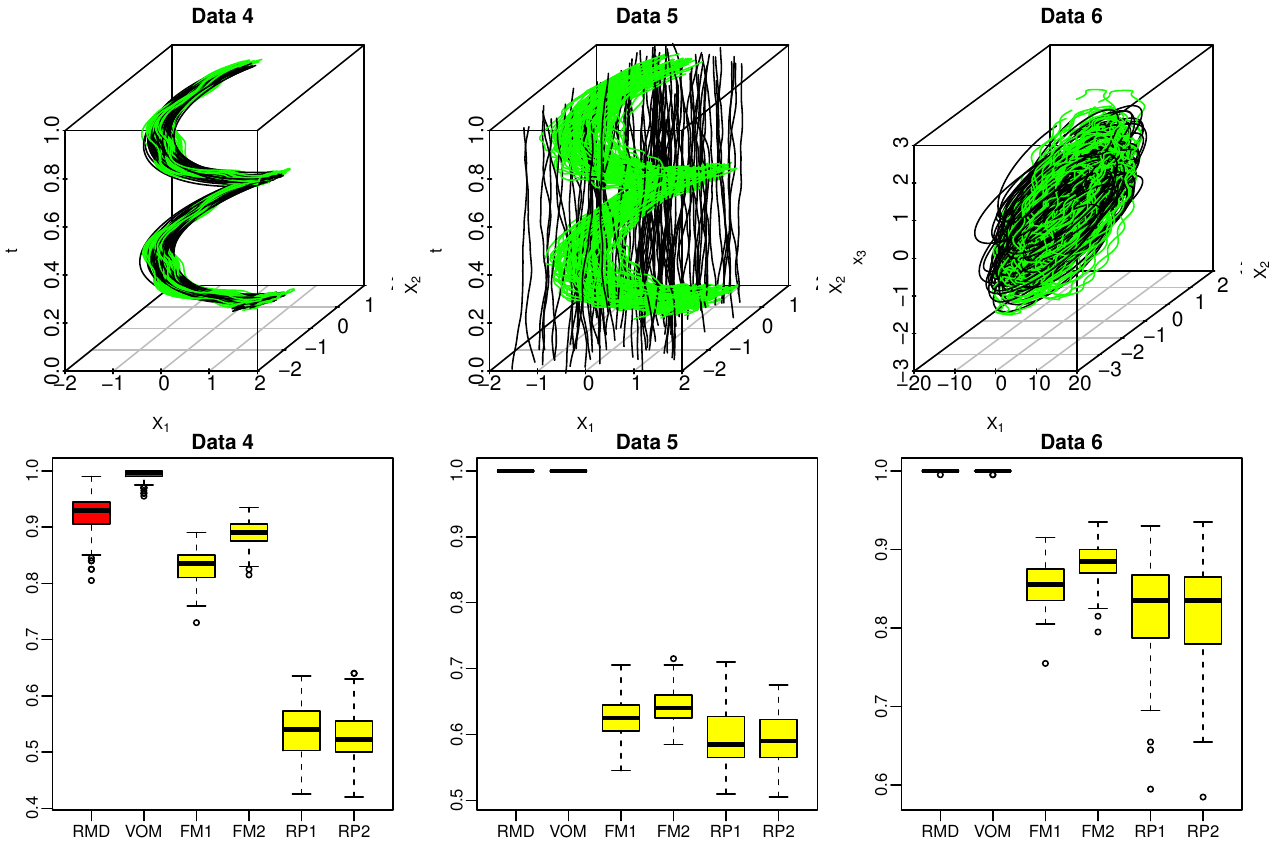}\\
\caption{Top panel: realizations of three multivariate functional data settings (Data 4, 5, 6). Bottom panel: correct classification rates of
our two proposed methods, RMD and VOM, and four existing methods, FM1, FM2, RP1, and RP2, for three settings based on 100 simulations.}
\label{plot_multivariate}
\end{center}
\end{figure}

Realizations of two classes of curves for each setting are illustrated in the top panel of Figure \ref{plot_multivariate}.
The functions were evaluated at 50 equidistant points from $[0,1]$, i.e. $t_i=i/50$.
We independently generated 200 samples from both classes of each data setting, randomly chose 100 of them as the training set, and treated the remaining 100 samples as the testing set.
We applied the six methods to the simulated data and calculated the correct classification rate for each method.
We repeated the above procedure 100 times and present the results in the bottom panel of Figure \ref{plot_multivariate}.
As illustrated, our proposed methods attain much higher $p_c$ than do the existing methods.
In particular, VOM has almost perfect classification results for the three settings. Sometimes the four existing methods provide results that are slightly better than results from completely random classification.
Data 5 is an example.
These simulation results again validate our claim that the proposed methods based on directional outlyingness are much more effective in distinguishing curve groups that vary by shape.

\begin{figure}[t!]
\begin{center}
\includegraphics[width=16.5cm,height=5.5cm]{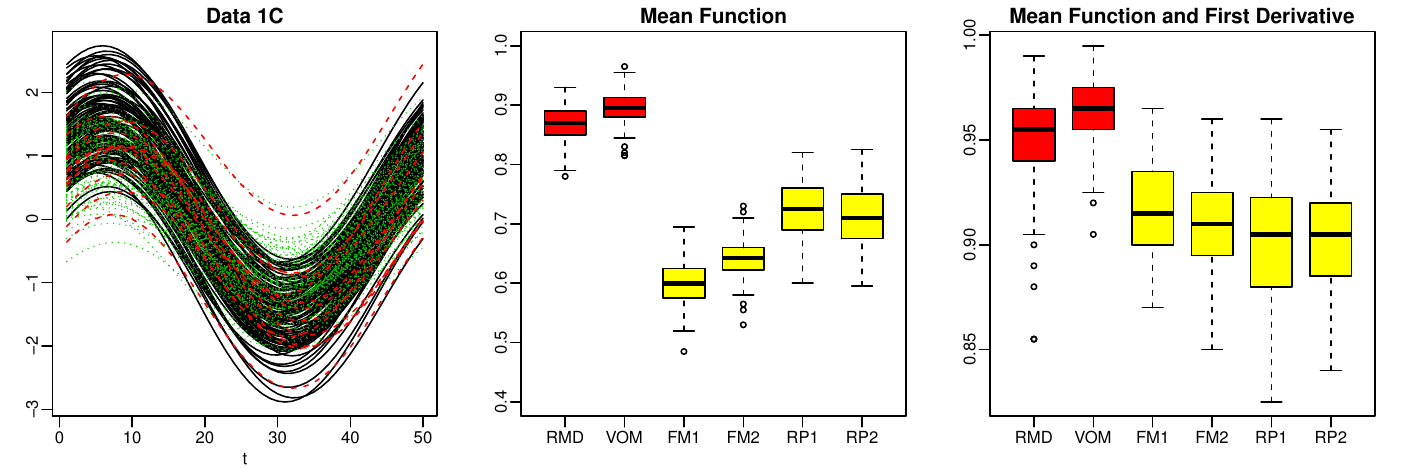}\\
\caption{Left plot: realizations of the setting of Data 1C with two classes. The long-dashed curves are the outliers contaminating Class 0. Middle plot: correct classification rates of the six methods using the mean curves.
Right plot: correct classification rates of the six methods using the combination of the mean curves and their first-order derivatives.}
\label{contaminated}
\end{center}
\end{figure}

Besides the non-contaminated settings, we also considered a contaminated setting.
\textbf{Data 1C}. Class 0: $X_0(t)=\{I_{(V\ge 0.1)}u_{01}+(1-I_{(V\ge 0.1)})u_{11}\}\sin(2\pi t)+u_{02}\cos(2 \pi t)+\varepsilon(t)$ and class 1: $X_1(t)=u_{11}\sin(2\pi t)+u_{12}\cos(2 \pi t)+\varepsilon(t)$, where $I_{A}$ is an indicator function: $I_{x}$ equals to $1$ if $x\in A$ and $0$ otherwise; $V$ was generated from $U(0,1)$. Class 0 was contaminated by outliers with a probability of $0.1$. \citet{sguera2014spatial} considered a similar setting.
The functions were evaluated at 50 equidistant points on $[0,1]$, i.e. $t_i=i/50$.
We independently generated 200 samples from both classes, randomly chose 100 of them as the training set, and treated the remaining 100 samples as the testing set.
We calculated the correct detection rates of the six methods based on the mean curves and the combination of the mean curves and their first-order derivatives, respectively.
The results as illustrated in Figure~\ref{contaminated}, are quite similar to the results from Data 1, suggesting that our proposed methods are robust to the presence of outliers.

\vskip 24pt
\section{Data Applications}
We evaluated our methods on two datasets: the first univariate and the second multivariate. Comparisons with existing methods are provided as well.

\vskip 12pt
\subsection{Phoneme Data}


We applied our methods to the benchmark phoneme dataset. Phoneme is a speech-recognition problem introduced by \citet{hastie1995penalized}. We obtained the data from the R package \emph{fds}.
The dataset comprises five phonemes extracted from the TIMIT database (TIMIT Acoustic-Phonetic Continuous Speech Corpus, NTIS, U.S. Department of Commerce).  
The phonemes are transcribed as follows: ``sh'' as in ``she'', ``dcl'' as in ``dark'', ``iy'' as the vowel in ``she'', ``aa'' as the vowel in ``dark'', and ``ao'' as the first vowel in ``water''. 
A log-periodogram was computed from each speech frame; this is one of several widely used methods for translating speech data into a form suitable for speech recognition.
For each log-periodogram, we considered the first 150 frequencies. 
In our study, we randomly selected 400 samples for each class and consequently, 2000 samples were considered in total. 
Ten samples from each class are illustrated in Figure S2 of the supplement. 
As shown, the five types of curves vary within the same range with different shapes.

\begin{figure}[t!]
\begin{center}
\includegraphics[width=16.5cm,height=8cm]{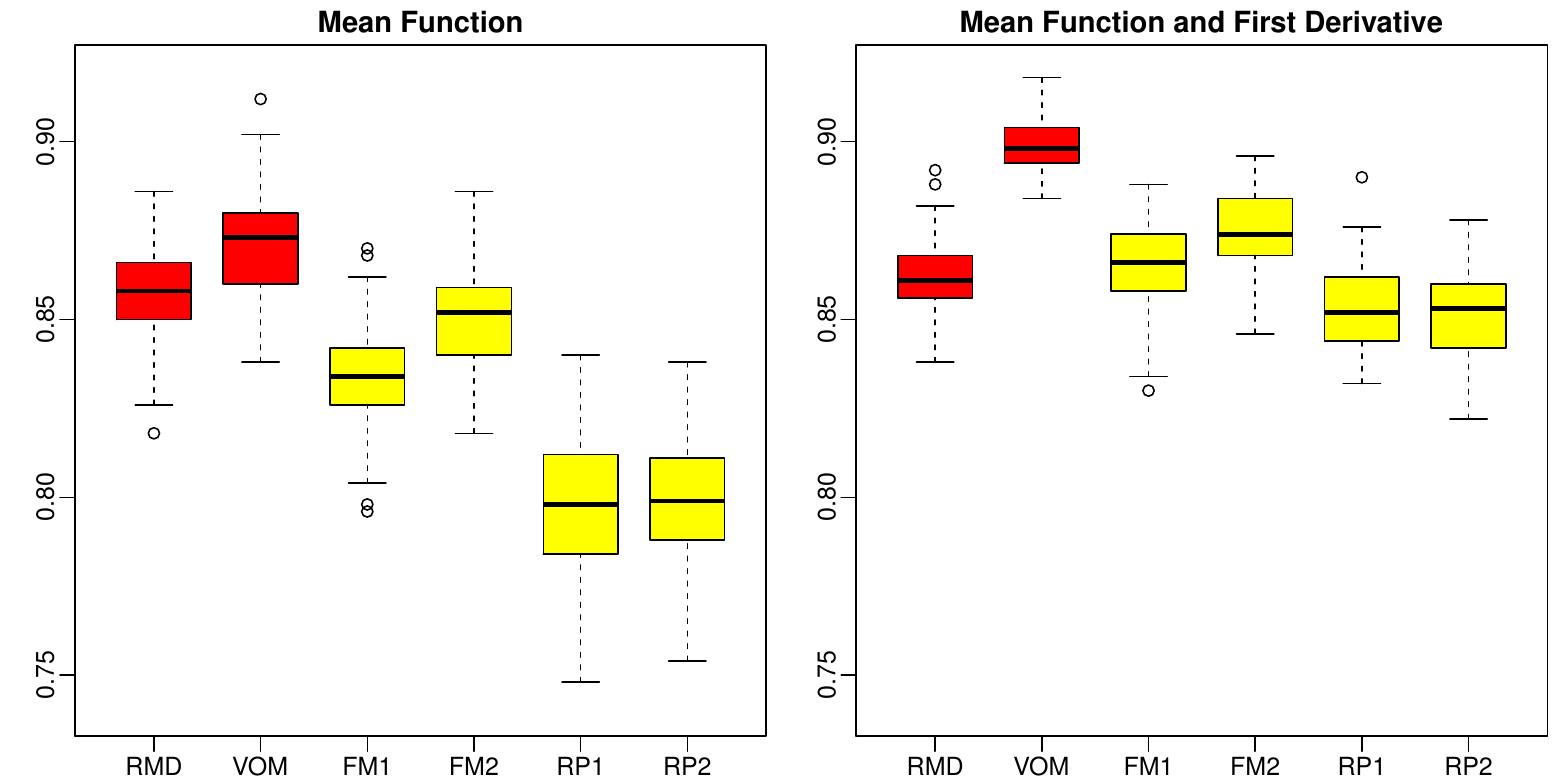}\\
\caption{Correct classification rates of our two proposed methods RMD and VOM, and four existing methods, FM1, FM2, RP1, and RP2, of the phoneme data. Left: results based on only raw data; right: results based on both raw data and their first-order derivatives.}
\label{plot_phoneme}
\end{center}
\end{figure}

We randomly selected 1500 samples as the training set (300 for each class) and treated the remaining 500 samples as the testing set (100 for each class).
We applied the six aforementioned methods in two ways: using only the raw data (univariate); using both raw data and their first-order derivatives (bivariate).
For each method, we calculated the correct classification rate and repeated this procedure 50 times.
The simulation results are presented in Figure \ref{plot_phoneme}. Based on the raw data, our methods perform better than the existing methods.
After taking their first derivatives into consideration, the performance of all methods except for RMD is improved significantly and VOM achieves the highest correct classification rate.

\vskip 12pt
\subsection{Gesture Data}
Gesture commands are widely used to interact with or control external devices, e.g., playing gesture-based games and controlling interactive screens.
The problem is how to recognize one observation accurately as a particular gesture.
Our second dataset includes gesture data comprising the eight simple gestures shown in Figure S3 of the supplement.
These gestures have been identified by a Nokia research study as preferred by users for interaction with home appliances.

\begin{figure}[!t]
	\begin{center}
		\includegraphics[width=16.5cm,height=6cm]{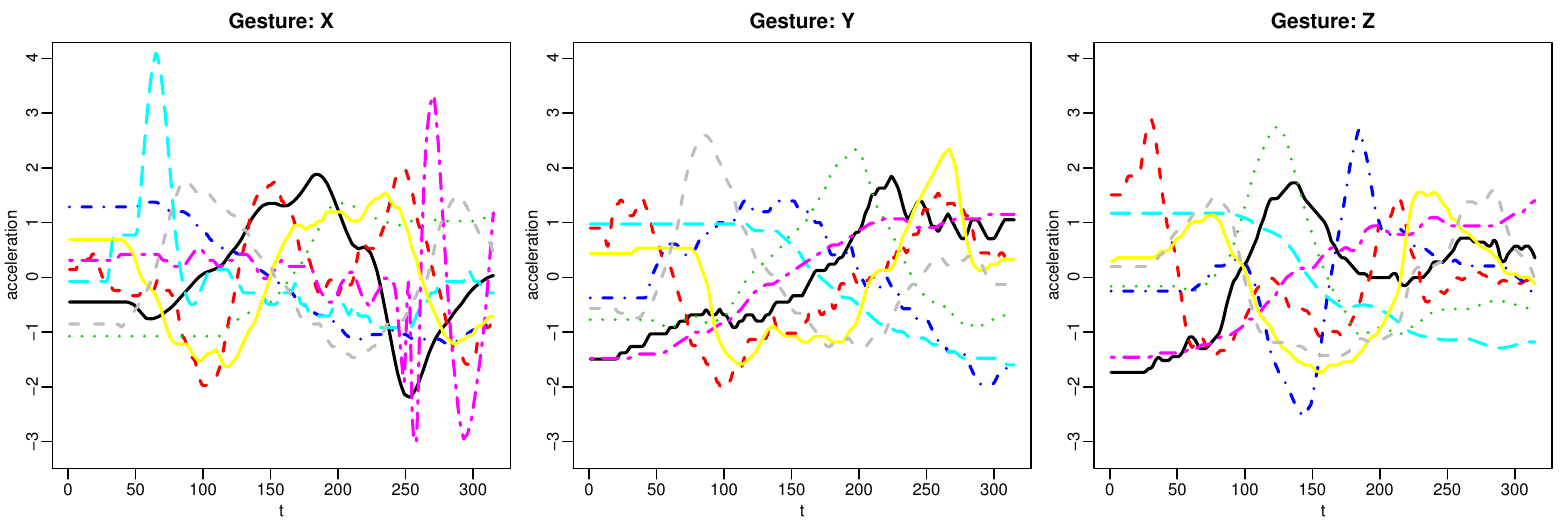}\\
		\caption{Left column: eight median curves of $X$-accelerations of eight gestures; middle column: eight median curves of $Y$-accelerations; right column: eight median curves of $Z$-accelerations.}
		\label{gesture_data_show}
	\end{center}
\end{figure}

\begin{figure}[!b]
\begin{center}
\includegraphics[width=16.5cm,height=16cm]{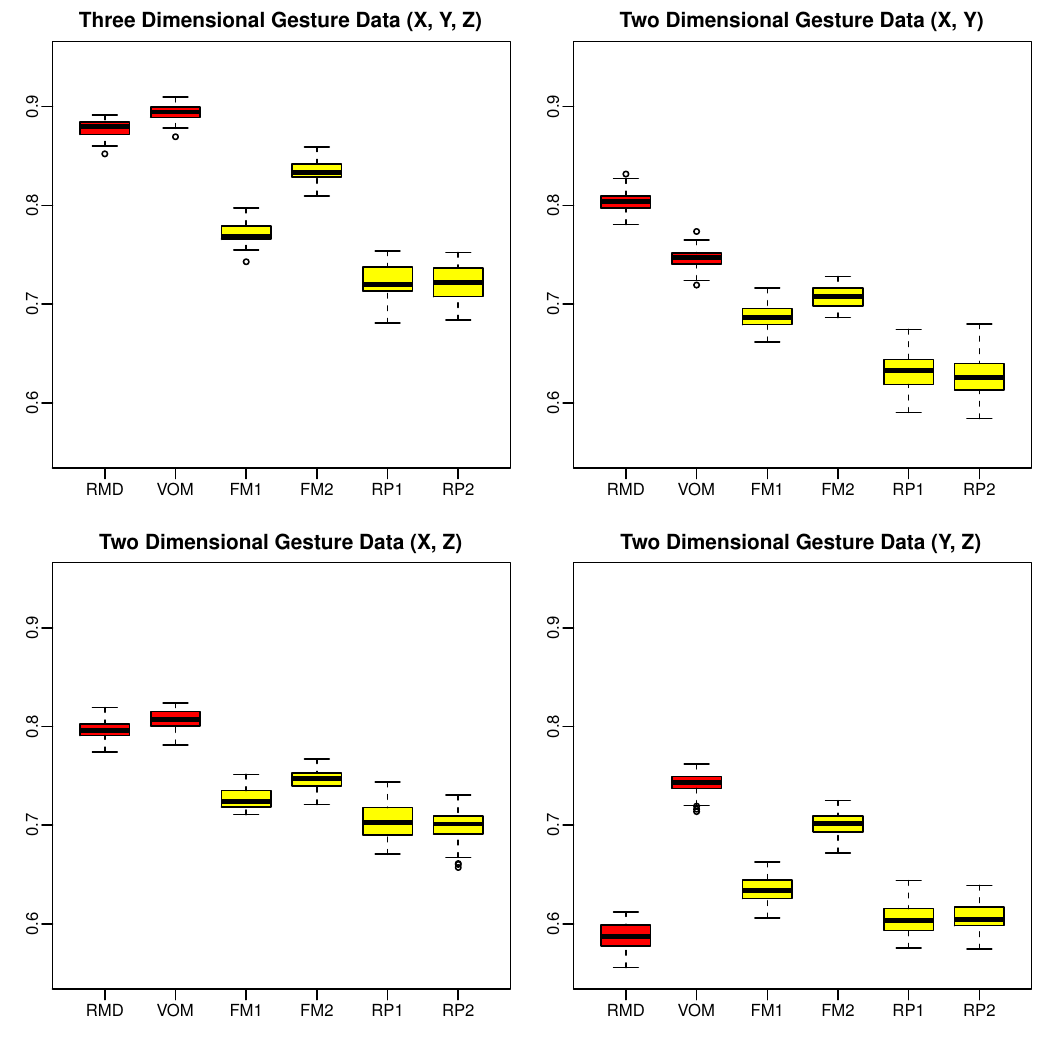}\\
\caption{Correct classification rates of our two proposed methods, RMD and VOM, and four existing methods, FM1, FM2, RP1, and RP2, of the gesture data. Top left: gesture data $(X,Y,Z)$; top right: gesture data $(X,Y)$;
bottom left: gesture data $(X,Z)$; bottom right: gesture data $(Y,Z)$.}
\label{plot_gesture_result}
\end{center}
\end{figure}

We downloaded this dataset from \citet{UCRArchive}. This dataset has been analyzed by \citet{shokoohi2015generalizing} with the dynamic time warping algorithm in a time series context. We used it to illustrate our functional data analysis approach. 
It includes 4,480 gestures: 560 for each type of action made by eight participants ten times per day during one week.
Each record contains accelerations on three orthogonal directions ($X$, $Y$ and $Z$), which means we need to classify three-dimensional curves. 
We found the median curve of acceleration for three directions of each gesture with the functional boxplot (\citet{sun2011functional}) as shown in Figure \ref{gesture_data_show}.
Generally, most of the acceleration curves oscillate between $-3$ and $3$.
We applied the six methods to the gesture data in four ways: combining all three components together, $(X,Y,Z)$, and selecting two components out of three, $(X,Y)$, $(X,Z)$, and $(Y,Z)$.
For each numerical study, we randomly selected 3200 samples as the training set (400 for each class) and treated the remaining 1280 samples as the testing set (160 for each class).
We repeated this procedure for 50 times and report the correct classification rates of each method in Figure~\ref{plot_gesture_result}.

In the four combinations, our proposed methods are always better than the four existing methods except for RMD of $(X,Z)$.
For three cases, VOM achieves the best performance among the six methods.
Overall, the correct classification rates improve as we raise the dimensions of the curves.
We define the marginal effect of component $X$ as the averaged difference between $p_c$ for $(X,Y,Z)$ and $(Y,Z)$.
This quantity measures how informative a component is for a classification task.
By comparing the plot of $(X,Y,Z)$ with the other three cases, we find that the marginal effect of $Y$ is the smallest.
This finding is consistent with the fact that the acceleration curves in direction $Y$ are more alike with each other. For example, the black and yellow curves in the middle graph of Figure \ref{gesture_data_show} are quite similar to the purple and red curves, respectively.
In contrast, the shapes of the acceleration curves in the other two directions differ, which leads to their higher marginal effects.
The gestures included in the dataset were mainly collected from the screens of smart phones, which means that the direction orthogonal to the screen is not as informative as the other two directions.

\vskip 24pt
\section{Discussion}

The proposed methods can be simply generalized to image or video data (\citet{genton2014surface}), where the support of functional data is two-dimensional.
We plan to investigate more general settings for both classifiers and data structures.
Rather than the constant weight function considered in the current paper, we believe that a weight function proportional to local variation could further improve our methods.
It is reasonable to put more weight on the time points where the curves differ a lot and less weight on those where the curves are quite alike.
For functional data observed at irregular or sparse time points (\citet{lopez2011depth}), we may fit the trajectories with a set of basis functions and then estimate depth of the discrete curves based on their continuous estimates.
The functional data within each group could be correlated in general data structures. An example is spatio-temporal precipitation (\citet{sun2012adjusted}).
Our methods need further modifications to account for the correlations between functional observations as well.

\section*{Acknowledgment}
The authors thank the editor, an associate editor, and the two referees for their constructive comments that led to a substantial improvement of the paper. The work of Wenlin Dai and Marc G. Genton was supported by King Abdullah University of Science and Technology (KAUST).

\section*{Supplementary Material}
Supplementary material includes an illustrative example of functional directional outlyingness framework, two figures of real data, and technical proofs for the theoretical results.

{
\vskip 24pt

\bibhang=1.7pc
\bibsep=2pt
\fontsize{9}{14pt plus.8pt minus .6pt}\selectfont
\renewcommand\bibname{\large \bf References}
\expandafter\ifx\csname
natexlab\endcsname\relax\def\natexlab#1{#1}\fi
\expandafter\ifx\csname url\endcsname\relax
\def\url#1{\texttt{#1}}\fi
\expandafter\ifx\csname urlprefix\endcsname\relax\def\urlprefix{URL}\fi

\bibliographystyle{chicago}
\bibliography{classification}
}

\end{document}